# Characterization of Acoustic Losses in Interdigitated VHF to mmWave Piezoelectric M/NEMS Resonators

Luca Colombo, Gabriel Giribaldi, Ryan Tetro, Jack Guida, Walter Gubinelli, Luca Spagnuolo,
Nicol Maietta, Siddhartha Ghosh, and Matteo Rinaldi
Institute for NanoSystems Innovation (NanoSI), Northeastern University, Boston, MA, USA

*Abstract*—This work reports on a technology-agnostic and frequency-independent methodology combining *a-priori* modeling, Finite Element Analysis (FEA), and experimental results for the characterization of acoustic losses in interdigitated piezoelectric micro- and nano-electromechanical (M/NEMS) resonators. The proposed approach models the mechanical quality factor ($Q_m$) and its dependency on piezoelectric ($Q_{piezo}$) and metal ($Q_{metal}$) acoustic losses, as a function of the mode of vibration dispersion characteristics. The model is finally experimentally validated by exploiting the intrinsic on-chip multifrequency manufacturability of interdigitated devices.

A broad range of available resonator technologies, frequencies, and piezoelectric materials are investigated for this purpose, including bulk X-cut Lithium Niobate (XLN) leaky surface acoustic wave resonators operating at Ultra High Frequency (UHF), thin film XLN Lamb Wave resonators spanning between Very High Frequency (VHF) and the $K_u$ band, and Aluminum Nitride (AlN) and scandium-doped AlN (ScAlN) cross-sectional lamé mode resonators ranging from the $K_u$ to $K_a$ band (mmWave).

## I. INTRODUCTION

MICROACOUSTIC piezoelectric resonators have been pivotal components for the implementation of modern Radio Frequency (RF) communication paradigms [1]. However, research continues to explore the capabilities of micro- and nanoelectromechanical systems (M/NEMS) for cutting-edge applications, particularly within the 5G FR-2 (mmWave) and anticipated FR-3 (X to $K_u$) bands [2]. In fact, as frequency increases, acoustic losses become more pronounced, hindering the implementation of microacoustic filters beyond 5 GHz. This challenge persists across all proposed technologies, either based on nano-scaled fundamental modes or overtones, and piezoelectric materials [3].

Over the years, several techniques to evaluate acoustic losses and their equivalent quality factor ($Q$) have been adopted. These methods include direct $Q$ extraction via acoustic delay lines of different lengths [4], cryogenic resonator characterization [5], or Brillouin scattering [6]. Each technique often requires dedicated equipment or comprehensive experimental studies with custom-designed structures. More recently, hybrid modeling based on Finite Element Analysis (FEA) and experimental data has been proposed to predict the impact of losses on the performance of Film and Overmoded Bulk Acoustic Resonators (FBARs/OBARs) [7].

Based on these findings, the present study introduces a data-driven and efficient method to characterize acoustic losses across various piezoelectric M/NEMS technologies excited via Interdigitated Transducers (IDTs). In the first section, the hybrid approach described in [7] is expanded to operate over a broad frequency range by leveraging the intrinsic lithographic multifrequency manufacturability and dispersive nature of IDT resonators. Subsequently, the framework for reconciling COMSOL® FEA and experimental data is presented. Then, the proposed characterization methodology is benchmarked against several readily available technologies, including X-cut Lithium Niobate (XLN) shear horizontal ($SH_0$) Leaky Surface Acoustic Wave (LSAW) resonators operating at Ultra High Frequency (UHF), XLN lateral ($S_0$) Lamb Wave Laterally Vibrating Resonators (LVRs) operating between Very High Frequency (VHF) and the $K_u$ band, and Aluminum Nitide (AlN) and 30%-doped Scandium Aluminum Nitride (ScAlN) Cross-Sectional Lamé Mode Resonators (CLMRs) operating between the $K_u$ and the $K_a$ or mmWave bands (Fig. 1a). In the third and final section, the paper presents remarks and suggestions for further enhancing the presented acoustic losses modeling.

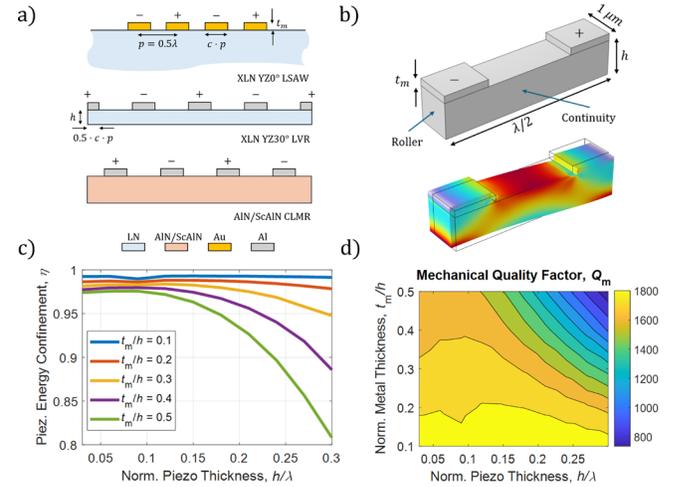

Fig. 1. a) Investigated resonator technologies: bulk X-cut YZ0° LN LSAWs [8], thin film X-cut YZ30° lithium niobate $S_0$ mode LVRs [5] [9], and thin film AlN [10] and 30% doped ScAlN CLMRs [11]; b) COMSOL® 2.5D model of a X-cut YZ170° $SH_0$ resonator and stress distribution at resonance; c) Piezoelectric energy confinement ($\eta$) as a function of the normalized film thickness ($h/\lambda$) for different normalized metal thicknesses ($t_m/h$); and d) Mechanical quality factor ($Q_m$) at resonance assuming constant $Q_{piezo}$ = 2000, $Q_{metal}$ = 200, and the $\eta$ distribution reported in Fig. 1c (Eq. 2). For the acoustic mode reported in this example, $Q_m$ severely degrades for thicker metal electrodes at higher frequencies.

## II. METHODOLOGY

The proposed methodology consists of a hybrid approach combining Finite Element Analysis (FEA) and experimental data to estimate the impact of acoustic losses in the piezoelectric ($Q_{piezo}$) and metal ($Q_{metal}$) domains on the mechanical quality factor at resonance ($Q_m$) of interdigitated piezoelectric resonators. In the proposed model, $Q_{piezo}$ represents all the acoustic losses within the piezoelectric medium and can be expressed according to Eq. 1:

$$\frac{1}{Q_{piezo}} = \frac{1}{Q_{p-p}} + \frac{1}{Q_{anchor}} + \frac{1}{Q_{air}} + \frac{1}{Q_{ni}} \quad (1)$$

where $Q_{p-p}$ represents phonon-phonon losses, $Q_{anchor}$ describes anchor's losses [12], $Q_{air}$ captures air damping [5], and $Q_{ni}$ models any other non-ideality, including fabrication. A similar expression for the acoustic losses in the metal can be granted by the addition of thermoelastic damping ($Q_{T\ ED}$) and electron-phonon losses ($Q_{e-p}$) [13]. For the sake of simplicity, all the terms of $Q_{piezo}$ beside $Q_{ni}$ are generally lumped into a constant value, while $Q_{metal}$ is modeled through a constant product between frequency and quality factor ($f \cdot Q$) according to literature [14]. Ultimately, the impact of $Q_{piezo}$ and $Q_{metal}$ on $Q_m$ is mediated through the piezoelectric energy confinement ($\eta$), according to Eq. 2 [10]:

$$Q_m = \frac{1}{\eta \cdot Q_{piezo}^{-1} + (1-\eta) \cdot Q_{metal}^{-1}} \quad (2)$$

More specifically, $\eta$ captures the ratio between the strain energy stored in the piezoelectric material and the total elastic energy present in the system. It can be estimated via COMSOL® 2.5 or 3D FEA, according to Eq. 3:

$$\eta = \frac{\iiint_{piezo} W(x,y,z)\, dx\, dy\, dz}{\iiint_{tot} W(x,y,z)\, dx\, dy\, dz} \quad (3)$$

where $W(x, y, z)$ represents the elastic strain energy density field [8]. The $\eta$ extraction procedure is exemplified for an X-cut YZ170° LN structural element in which the fundamental shear horizontal mode ($SH_0$) is excited by a couple of aluminum half-electrodes covering 50% of the top surface (Fig. 1b) [15]. The structural element possesses a width equal to half of the desired acoustic wavelength ($\lambda/2$), a height equal to the piezoelectric film thickness ($h$), and depth equal to 1 μm. The proper boundary conditions are chosen to emulate an infinite plate, while a mechanical isotropic damping lower than 1/100 is set to accurately evaluate the integral in Eq. 3. As highlighted in Fig. 1c, $\eta$ drastically decreases for thicker electrodes ($t_m$) and larger normalized piezoelectric film thicknesses ($h/\lambda$) due to the presence of a larger strain field in the metal. By assuming constant $Q_{piezo} = 2,000$ and $Q_{metal} = 200$, it is possible to appreciate the sharp $Q_m$ decrease associated with larger $\eta$ (Fig. 1d). This approach is generally valid since metals are inherently lossier than ceramic materials [14].

## III. EXPERIMENTAL VALIDATION

To validate the approach described in Section II, the generic $Q_m$ model (Eq. 2) is compared to the measured quality factor at resonance ($Q_{3dB}$), properly de-embedded of the ohmic losses [5] [13]. Differently from FBARs, interdigitated resonator technologies offer intrinsic on-chip multi-frequency manufacturability, allowing the synthesis of devices covering a large portion of the dispersion range [16]. By correctly guessing the distribution of $Q_{piezo}$ and $Q_{metal}$ as a function of $\lambda$ and coupling those functions through $\eta$, it is possible to characterize the nature of the acoustic losses for several classes of devices. Two $Q_m$ models are developed for each technology as a proof of concept. The former generally assumes a constant $Q_{piezo}$ which is frequency-independent, while the latter modulates the piezoelectric losses through a finite non-ideality quality factor ($Q_{ni}$). In both cases, $f \cdot Q_{metal}$ is assumed constant [14].

### A. X-cut LN Leaky Surface Acoustic Wave Resonators

X-cut YZ170° $SH_0$ Leaky SAW resonators [8] with gold electrodes and operating at UHF (450 to 600 MHz) are used to validate the model on bulk substrates (Fig. 2a-c). Unlike thin film technologies, LSAWs dispersion characteristics are

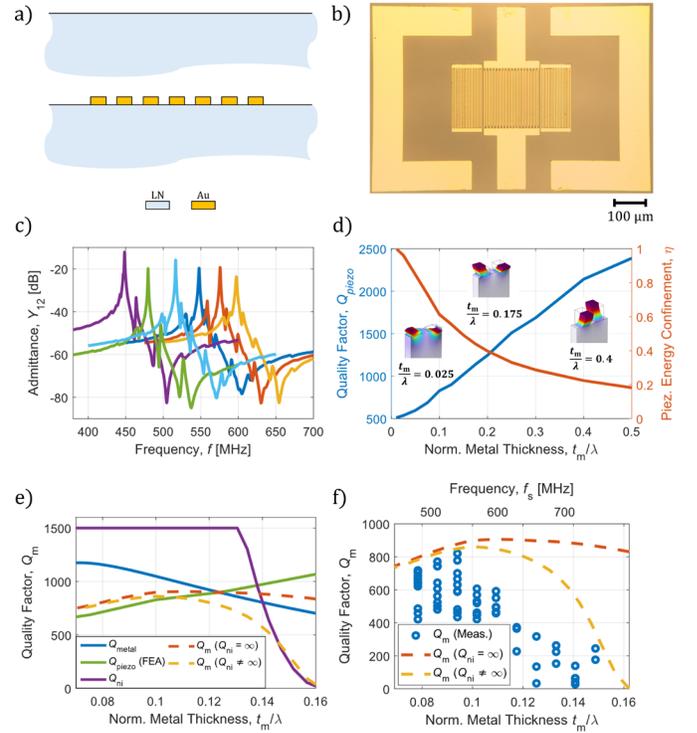

Fig. 2. a) XLN LSAW process flow, consisting of a single lithographic step followed by the sputtering and lift-off of 300 nm of gold (Au) [8]; b) Microscope image of a fabricated XLN LSAW operating around 450 MHz; c) Admittance responses of fabricated LSAWs operating between 450 and 600 MHz; d) COMSOL® 2.5D FEA simulated piezoelectric confinement ($\eta$) and leakage quality factor according to FEA ($Q_{piezo}$) as a function of the normalized electrode thickness ($t_m/\lambda$); e) Modeled quality factors ($Q$) distributions as a function of $t_m/\lambda$. A $f \cdot Q_{metal}$ product of 0.5 THz is assumed for Au; and f) Modeled mechanical quality factor ($Q_m$) versus measured data as a function of $t_m/\lambda$ and frequency.

## C. ScAlN and AlN Cross-sectional Lamé Mode Resonators

Finally, 30% doped ScAlN and AlN CLMRs operating in the K band are used to validate the model for sputtered, cross-field excited, thin film resonators. The devices are fabricated on 150 nm and 100 nm in-house sputtered thin films on high resistivity silicon and share fabrication process and design geometries (Fig. 5a-b) [10] [11]. According to the presented methodology, ScAlN CLMRs exhibit a strong $Q_m$ decrease over frequency due to a combined reduction of $Q_{metal}$ and $Q_{ni}$ (Fig. 5e-f), while mmWave AlN CLMRs appear to be mainly limited by acoustic losses in the metallic layer up to 30 GHz (Fig. 6e-f).

## IV. CONCLUSION

This work presents a technology-agnostic and frequency-independent methodology for the characterization of acoustic losses in interdigitated resonators, which combines Finite Element Analysis and experimental data. The approach is successfully validated on various piezoelectric materials, modes of vibrations, and frequencies ranging from VHF to mmWave. Currently, the model adopts generic $Q_{piezo}$, $Q_{metal}$, and $Q_{ni}$ functions to fit $Q_m$ to the envelope of the measured $Q_m$. Further developments of the model will focus on an improved description of those functions, either from theoretical *a-priori* modeling or through experimental verification.

## V. ACKNOWLEDGEMENT

This work was supported by the Defense Advanced Research Project Agency (DARPA) COFFEE project under Contract HR001122C0088 and developed in collaboration with Raytheon Missiles & Defense and Naval Research Lab (NRL), by the U.S. Army Combat Capabilities Development Command Research Laboratory (ARL DEVCOM), and Rogers Corporation. The authors would also like to thank Northeastern University Kostas Cleanroom and Harvard CNS staff.

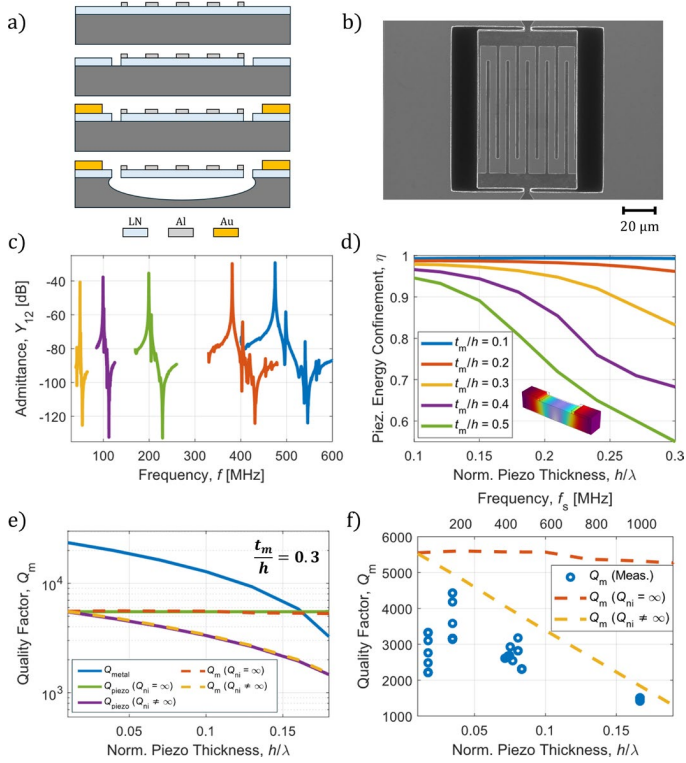

Fig. 3. a) XLN $S_0$ LVR process flow, consisting of aluminum (Al) top electrode patterning and lift-off, trenches definition via deep reactive ion etching (DRIE), gold (Au) pad patterning and lift-off, and xenon difluoride (XeF$_2$) isotropic release. 300 nm of Al and direct lithography are used for the MHz-range LVRs, while 50 nm of Al and e-beam lithography are used for the GHz-range devices [17] [9]; b) Scanning Electron Microscope (SEM) image of a fabricated MHz-range LVR; c) Admittance responses of fabricated $S_0$ LVRs operatng between 50 and 550 MHz; d) COMSOL® 2.5D FEA simulated piezoelectric confinement ($\eta$) as a function of the normalized film thickness ($h/\lambda$) and different $t_m/h$; e) Modeled quality factors ($Q$) distributions as a function of $h/\lambda$. A $f \cdot Q_{metal}$ product of 1.3 THz is assumed for Al; and f) Modeled mechanical quality factor ($Q_m$) versus measured data as a function of $h/\lambda$ and frequency.

expressed as a function of metal electrode thickness ($t_m$) over $\lambda$. Furthermore, $Q_{piezo}$ is strongly impacted by anchor losses due to substrate leakage and cannot be assumed constant over $t_m/\lambda$ (Fig. 2d). As reported in Fig. 2e-f, the losses model correctly predicts the maximum $Q_m$ achievable by the investigated technology, hinting to substrate acoustic leakage for $t_m/\lambda$ lower than 0.1 and to $Q_{ni}$ for higher frequencies.

## B. X-cut LN $S_0$ Lamb Wave Resonators

X-cut YZ30° $S_0$ LVRs with aluminum electrodes are used to validate the model for thin-film, single-crystal substrates excited via in-line electric fields. Two sets of devices fabricated on 1 µm and 100 nm XLN on high resistivity silicon and operating between 50 MHz and 16 GHz are investigated (Fig. 3a-c and Fig. 4a-b, respectively) [5] [9]. According to the proposed model, 1 µm LVRs spanning the sub-GHz range do not show $Q_m$ degradation due to metal losses, being rather limited by non-idealities (Fig. 3e-f). Differently, 100 nm thick devices ranging from the S to the K$_u$ band exhibit larger losses due to increased $\eta$ (Fig. 4c-d).

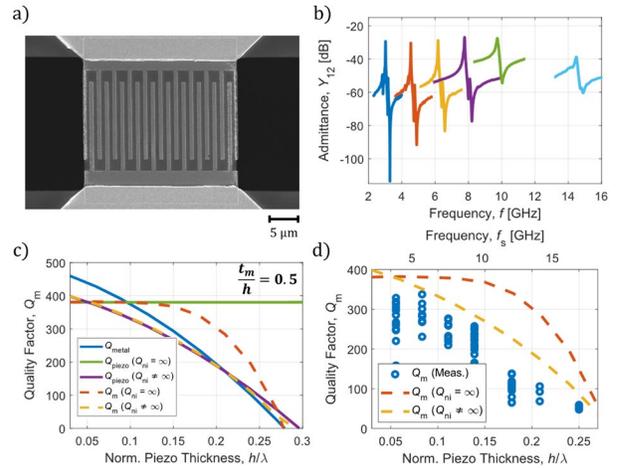

Fig. 4. a) Scanning Electron Microscope (SEM) image of a fabricated GHz-range LVR; b) Admittance responses of fabricated $S_0$ LVRs operating between 2 and 16 GHz; c) Modeled quality factors ($Q$) distributions as a function of the normalized film thickness ($h/\lambda$); and d) Modeled mechanical quality factor ($Q_m$) versus measured data as a function of $h/\lambda$ and frequency.

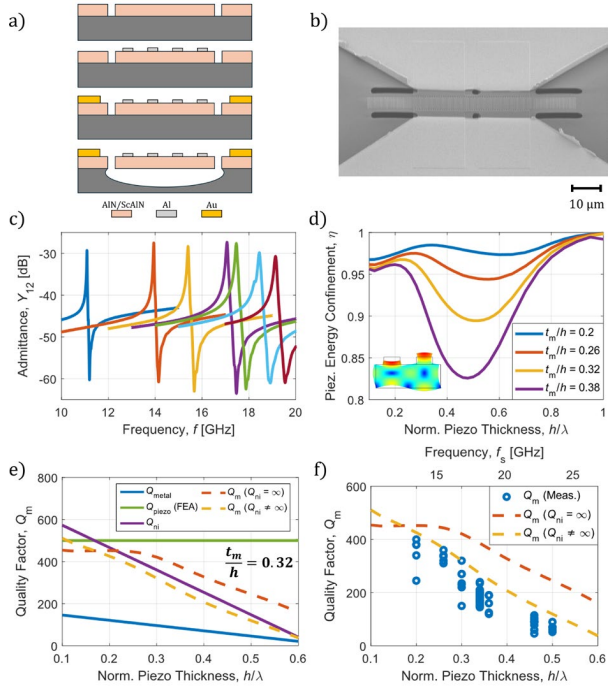

Fig. 5. a) CLMR process flow, consisting of thin film reactive sputter deposition, trench definition via ion milling, aluminum (Al) patterning and lift-off, gold (Au) pad patterning and lift-off, and xenon difluoride (XeF$_2$) isotropic release. 50 nm and 30 nm of Al are used for ScAlN and AlN CLMRs, respectively; b) Scanning Electron Microscope (SEM) image of a fabricated CLMR; c) Admittance responses of fabricated ScAlN CLMRs operating between 11 and 19 GHz; d) COMSOL® 2D FEA simulated piezoelectric confinement ($\eta$) as a function of the normalized film thickness ($h/\lambda$) and different $t_m/h$; e) Modeled quality factors ($Q$) distributions as a function of $h/\lambda$. A $f \cdot Q_{metal}$ product of 1.3 THz is assumed for Al; and f) Modeled mechanical quality factor ($Q_m$) versus measured data as a function of $h/\lambda$ and frequency.

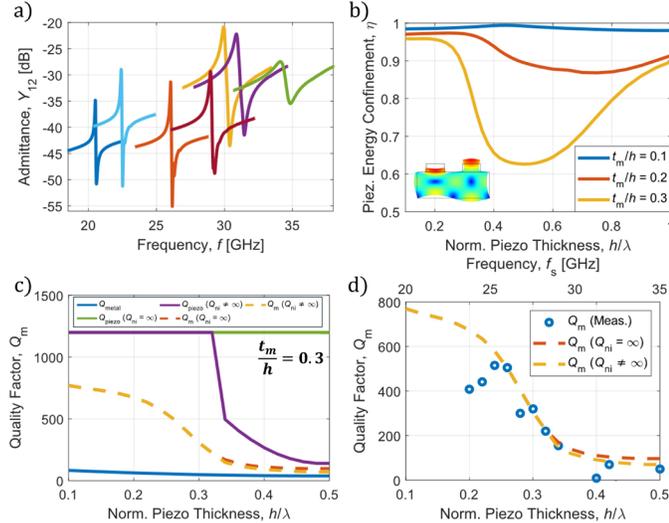

Fig. 6. a) Admittance responses of fabricated AlN CLMRs operating between 20 and 35 GHz; d) COMSOL® 2D FEA simulated piezoelectric confinement ($\eta$) as a function of the normalized film thickness ($h/\lambda$) and different $t_m/h$; e) Modeled quality factors ($Q$) distributions as a function of $h/\lambda$. A $f \cdot Q_{metal}$ product of 1.3 THz is assumed for Al; and f) Modeled mechanical quality factor ($Q_m$) versus measured data as a function of $h/\lambda$ and frequency.